\documentclass[journal=nalefd,manuscript=letter,layout=onecolumn]{achemso}

\usepackage{graphicx}
\usepackage{dcolumn}
\usepackage{bm}
\usepackage{gensymb}
\usepackage{siunitx}
\usepackage{amsmath}
\usepackage{amssymb}

\author{Riko Kiessling}
 \affiliation{Fritz-Haber-Institut der Max-Planck-Gesellschaft, Faradayweg 4-6,14195 Berlin, Germany}
\author{Yujin Tong}
 \affiliation{Fritz-Haber-Institut der Max-Planck-Gesellschaft, Faradayweg 4-6,14195 Berlin, Germany}
\author{Alexander J. Giles}
 \affiliation{US Naval Research Laboratory, 4555 Overlook Avenue SW, Washington DC 20375, USA}
\author{Sandy Gewinner}
 \affiliation{Fritz-Haber-Institut der Max-Planck-Gesellschaft, Faradayweg 4-6,14195 Berlin, Germany}
\author{Wieland Sch\"ollkopf}
 \affiliation{Fritz-Haber-Institut der Max-Planck-Gesellschaft, Faradayweg 4-6,14195 Berlin, Germany}
\author{Joshua D. Caldwell}
 \affiliation{Department of Mechanical Engineering, Vanderbilt University, 2400 Highland Ave, Nashville, TN 37212, USA} 
 \alsoaffiliation{US Naval Research Laboratory, 4555 Overlook Avenue SW, Washington DC 20375, USA}
\author{Martin Wolf}
 \affiliation{Fritz-Haber-Institut der Max-Planck-Gesellschaft, Faradayweg 4-6,14195 Berlin, Germany}
\author{Alexander Paarmann}
 \affiliation{Fritz-Haber-Institut der Max-Planck-Gesellschaft, Faradayweg 4-6,14195 Berlin, Germany}
 \email{alexander.paarmann@fhi-berlin.mpg.de}

\title{Surface phonon polariton resonance imaging using long-wave infrared-visible sum-frequency generation microscopy}

\abbreviations{LWIR,IR,SFG,VIS,s-SNOM,PTIR,SPhP}
\begin{document}

\begin{tocentry}
\includegraphics{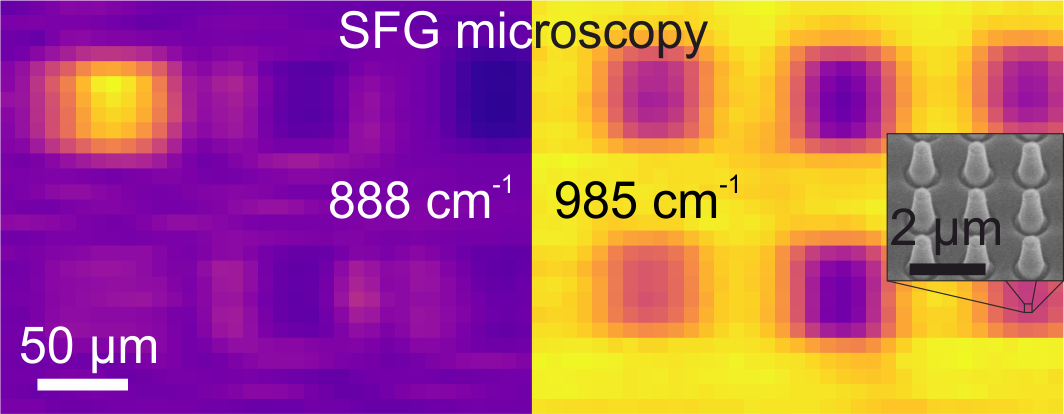}

%
%
%

\end{tocentry}

\begin{abstract}
We experimentally demonstrate long-wave infrared-visible sum-frequency generation microscopy for imaging polaritonic resonances of infrared (IR) nanophotonic structures. This nonlinear-optical approach provides direct access to the resonant field enhancement of the polaritonic near fields, while the spatial resolution is limited by the wavelength of the visible sum-frequency signal. As a proof-of-concept, we here study periodic arrays of subdiffractional nanostructures made of 4H-silicon carbide supporting localized surface phonon polaritons. By spatially scanning tightly focused incident beams, we observe excellent sensitivity of the sum-frequency signal to the resonant polaritonic field enhancement, with a much improved spatial resolution determined by visible laser focal size. However, we report that the tight focusing can also induce sample damage, ultimately limiting the achievable resolution with the scanning probe method. As a perspective approach towards overcoming this limitation, we discuss the concept of using wide-field sum-frequency generation microscopy as a universal experimental tool that would offer long-wave IR super-resolution microscopy with spatial resolution far below the IR diffraction limit.
\end{abstract}


Nanophotonics relies on the subdiffractional localization of light  which is typically achieved using large-momentum surface polariton modes.\cite{Maier2007,Caldwell2014a} This has been  demonstrated for a large variety of structures, using both plasmon polaritons in metallic nanoantennas\cite{Schuller2010,Giannini2011,Akselrod2014} as well as phonon polaritons in subdiffractional polar dielectric nanostructures,\cite{Urzhumov2007,Schuller2009,Caldwell2013a,Wang2013a,Caldwell2014,Feng2015,Autore2017,Wang2017b} emerging in the visible (VIS) and long-wave infrared (LWIR) spectral region, respectively. Experimental verification of the resulting light localization, however, is inherently challenging since the spatial resolution in optical microscopes is dictated by the diffraction limit. Specifically in the field of LWIR nanophotonics, this paradigm has been very successfully resolved by using scattering-type scanning near-field optical microscopy (s-SNOM)\cite{Zenhausern1994,Knoll1999} and photothermal induced resonance (PTIR) microscopy,\cite{Lahiri2013,Centrone2015} where the spatial resolution is limited by the nanoscopic size of the metallic probe tip rather than the imaging wavelength.\cite{Wang2013a,Dai2014,Wang2017b,Hillenbrand2002,Woessner2015,Giles2016,Huber2016,Giles2017,Li2018,Ma2018,Ambrosio2017,Brown2018,Khatib2018} Further approaches implementing hyperlens concepts may provide additional imaging methods for overcoming the challenge for the diffraction limit as well.\cite{Fang2005,Li2015,Dai2015a}

In biology and chemistry, however, the traditional way to achieve image resolution below the diffraction limit has been to harvest nonlinear-optical effects, such as stimulated emission depletion.\cite{Huang2009} Infrared (IR) super resolution imaging, on the other hand, has used nonlinear techniques such as IR-VIS sum-frequency generation\cite{Shen2016,Hoffmann2002,Cimatu2006,Raghunathan2011,Wang2017} (SFG) or coherent anti-stokes Raman scattering\cite{Zumbusch1999,Chung2013} microscopy, providing surface-specific and bulk vibrational contrast, respectively. For both approaches, the improved spatial resolution is derived simply from the short wavelength of the visible light that probes, through the nonlinear-optical process, the IR response. Additionally, the tensorial nature of the nonlinear susceptibility $\chi^{(2)}$ responsible for SFG allows for the extraction of the local chemical and structural information through careful analysis of signals from different polarization conditions.\cite{Sovago2009a} For materials with broken inversion symmetry, IR-VIS SFG spectroscopy reports on the bulk phonon resonances, weighted by local field effects.\cite{Liu2008} 

It is highly intriguing to transfer these concepts to the field of LWIR nanophotonics. Through the nonlinear-optical interaction, LWIR-VIS SFG spectroscopy would provide direct access to optical field enhancements associated with nanophotonic resonances, similar to LWIR second harmonic generation (SHG) spectroscopy.\cite{Razdolski2016,Passler2017,Razdolski2018,Ratchford2018} In contrast to SHG, however, the spatial resolution of IR-VIS SFG microscopy is ultimately limited by the SFG wavelength, only, allowing for microscopy well below the IR diffraction limit. Thus far, such studies have been hindered by the scarcity of high-intensity laser sources in the LWIR spectral region relevant for nanophotonics based on surface phonon polaritons (SPhPs).

In our work, we extend the IR wavelength range of IR-VIS SFG microscopy to the LWIR beyond 10~$\mu$m. As a proof-of-concept, here we study localized SPhPs in subdiffractional nanostructures made from 4H-Silicon Carbide (4H-SiC). We show that the SFG signals report on the optical field enhancements associated with the SPhP resonances. Using scanning probe SFG microscopy with tightly focused beams, we demonstrate improved spatial resolution and imaging contrast in comparison to the linear IR reflectance.

\begin{figure*}[bth!]
\includegraphics[width=.9\linewidth]{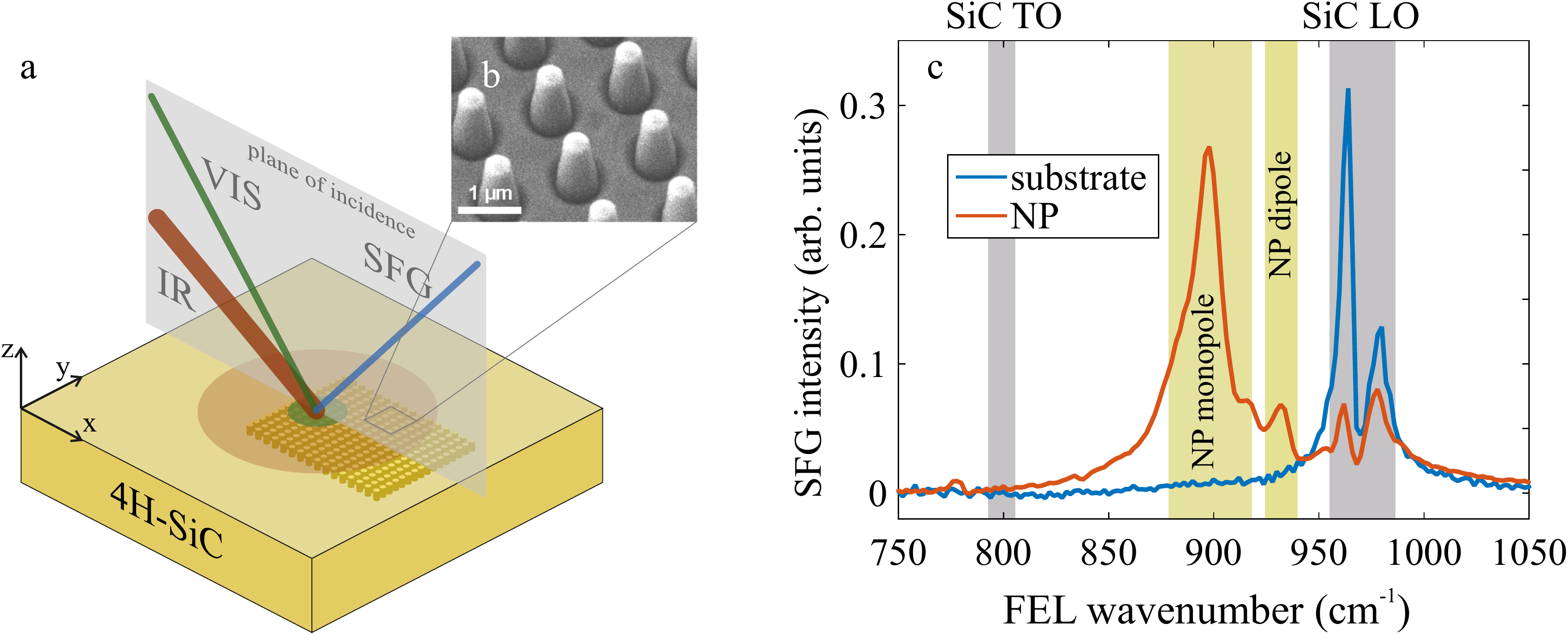}
  \caption{\textbf{Concept of LWIR-VIS SFG micro-spectroscopy of IR nanophotonic structures} a) Scheme of our experimental approach. A tunable, resonant IR laser and a nonresonant VIS laser are focused onto the nanostructure, generating a signal at the sum-frequency. The red and green shaded areas illustrate the respective incoming laser spots. The SFG is emitted from the overlapping region, resulting in improved achievable spatial resolution of the SFG compared to the IR reflectance. b) Electron microscope image of SiC nanopillars. c) SFG spectra of the 4H-SiC substrate (blue) and a nanopillar array (red). The grey shaded areas mark the position of the optical phonon resonances of SiC, while the yellow shaded areas highlight the SPhP resonances of the nanopillars.} 
  \label{fig1}
\end{figure*}

The experimental scheme is outlined in Fig.~\ref{fig1}a. VIS and IR laser pulses are spatially and temporally overlapped on the sample, and the generated sum-frequency intensity is detected. The IR free-electron laser (FEL) installed at the Fritz Haber Institute provides high-power, narrow-band (bandwidth of $\sim 0.3$\%), and wavelength-tunable mid-IR and LWIR light, and is described in detail elsewhere.\cite{Schollkopf2015} Notably, for SFG we operate the FEL at reduced electron micro-bunch repetition rate of 55.5~MHz, matched to the repetition rate of a synchronized tabletop laser which provides the VIS up-conversion beam. The latter is generated by frequency-doubling of the 1050~nm fiber oscillator output (OneFive, 100~fs pulse duration). Synchronization of the FEL micro-pulses and the VIS laser pulses is achieved by a phase-lock-loop electronics operating at 2.998 GHz, using the respective RF frequency reference from the accelerator system and the 54$^{th}$ harmonic of fiber-laser timing output, see ref.~\cite{Kiessling2018} for details of the timing and synchronization infrastructure. 

The IR and VIS (525~nm) laser beams are focused onto the sample with angles of incidence of 55 and 30 deg, respectively. The SFG light emitted in reflection is spectrally separated from the almost collinear reflected VIS beam by spectral edge filters and detected with a photo multiplier (Hamamatsu). SFG spectra are acquired by spectral scanning of the FEL wavenumber from 750 to 1050~cm$^{-1}$ using the motorized undulator gap. SFG microscopy is implemented by spatial scanning of the sample using a motorized three-axis translation stage at fixed FEL wavelength and tightly focused beams.

The SiC nanopillar samples were produced by e-beam lithography and dry etching of a semi-insulating 4H-SiC substrate, the details of the fabrication procedure are found elsewhere.\cite{Chen2014} Specifically, we fabricated multiple 50~$\mu$m$\,\times\,$50~$\mu$m sized arrays spaced 50$\mu$m apart, with varied nanopillar diameter and period ranging from 200 to 600~nm and 700 to 1500~nm, respectively; an electron microscope image of a representative arrays is shown in Fig. 1b). Localized SPhPs in these structures emerge within the Reststrahlen band of SiC between the transversal optical (TO) and longitudinal optical (LO) phonon frequencies at 797~cm$^{-1}$ and 964~cm$^{-1}$, respectively, where the dielectric permittivity of SiC is negative.\cite{Caldwell2013a,Chen2014,Caldwell2014a,Razdolski2016} 

A representative comparison of the SFG spectrum of one such nanopillar array with the response from the bare 4H-SiC substrate within the Reststrahlen spectral range are provided in Fig.~\ref{fig1}c. Both spectra where acquired in SSP polarization configuration (s-polarized SFG and VIS beams, p-polarized IR beam) with IR and VIS focal sizes matched to the array size. Conveniently, in this configuration the signal arises from a single $\chi^{(2)}$ tensor component.\cite{Paarmann2016} The SFG intensity for this specific configuration at frequency $\omega_{SFG} = \omega_{VIS} + \omega_{IR}$ is given by:\cite{Shen2016}
\begin{equation}
    I_{SSP}^{SFG} \propto | L_{yy}^{SFG}\chi^{(2)}_{yyz} L^{VIS}_{yy}E^{VIS}_yL^{IR}_{zz}E^{IR}_z|^2/\Delta k^2,
    \label{eq:SFG}
\end{equation}
where $E_m^{IR(VIS)}$ are the incoming field components $(m=y,z)$ and $L_{mm}^{IR(VIS,SFG)}$ the Fresnel transmission factors of the IR(VIS,SFG) beam. $\chi_{ijk}^{(2)}$ is the second order susceptibility tensor. Due to the broken inversion symmetry of 4H-SiC (point-group symmetry 6mm), a sizable bulk second order nonlinear signal arises. For the substrate spectrum, the effective source volume is determined by the wave vector mismatch $\Delta_k = |\mathbf{k}_{SFG} - \mathbf{k}_{VIS} - \mathbf{k}_{IR}|$ inside the SiC crystal,\cite{Paarmann2015,Paarmann2016} while for the nanopillar resonances this is dictated by the localized mode volume.\cite{Caldwell2013a,Chen2014}

In the substrate spectrum, a double peak feature is observed in the region of the LO phonon at $\omega_{IR}\sim 960-990$~cm$^{-1}$, which is due to resonances in the normal-to-surface Fresnel transmission factor $L_{zz}$:\cite{Paarmann2016}
\begin{equation}
    L_{zz}(\omega) = \frac{\epsilon_{\perp}(\omega)}{\epsilon_\parallel(\omega)}\frac{2k_z^{air}(\omega)}{\epsilon_\perp(\omega) k_z^{air}(\omega)+k_{z,e}^{SiC}(\omega).}
    \label{eq:fresnel}
\end{equation}
Here, $\epsilon_{\perp(\parallel)}(\omega)$ are the ordinary (extraordinary) dielectric function of c-cut uniaxial 4H-SiC, and $k_z^{air}(k_{z,e}^{SiC})$ the normal-to-surface wave vector in air (in 4H-SiC).\cite{Paarmann2016} The resulting enhancement of the local optical fields inside the SiC crystal $E_z^{SiC} = L_{zz}E_z^{air}$ and the small wave vector mismatch $\Delta k$ are largely responsible for the enhancement of the SFG signal, since all other quantities in Eq.~\ref{eq:SFG} do not exhibit a significant dispersion in this frequency range. Therefore, the two peaks in the SFG spectra can be assigned to the resonant denominators of the two factors in Eq.~\ref{eq:fresnel}, respectively, very similar to what was previously observed in SHG spectroscopy,\cite{Paarmann2016,Paarmann2015,Razdolski2016} see Supporting Information for a theoretical description of the substrate SFG spectrum.

Moving the nanophotonic structures into the laser foci drastically changes the SFG spectra. First, the SFG resonances in the LO phonon region are strongly suppressed on the structured surface, due to modification of the effective Fresnel factors of the VIS and SFG beam in Eq.~\ref{eq:SFG}. Namely, the period of the nanopillar array being similar to the VIS wavelength results in diffraction of the VIS and SFG beams. In consequence, the SFG signal experiences an approximately five-fold suppression. More important, new resonances emerge due to the IR field enhancement associated with the localized SPhPs.\cite{Caldwell2013a,Chen2014,Razdolski2018,Razdolski2016} Specifically, the strongest SPhP resonance at $\omega_{IR}\sim890$~cm$^{-1}$ can be assigned to the monopolar mode featuring pronounced normal-to-surface field enhancements, while the peak at $\omega_{IR}\sim935$~cm$^{-1}$ corresponds to excitation of a dipolar mode. The latter predominantly results in an in-plane enhancement of the IR fields\cite{Chen2014,Razdolski2016} with only mild enhancement of the out-of-plane fields that are probed by SFG in this polarization configuration. 

\begin{figure*}[bth!]
\includegraphics[width=1\linewidth]{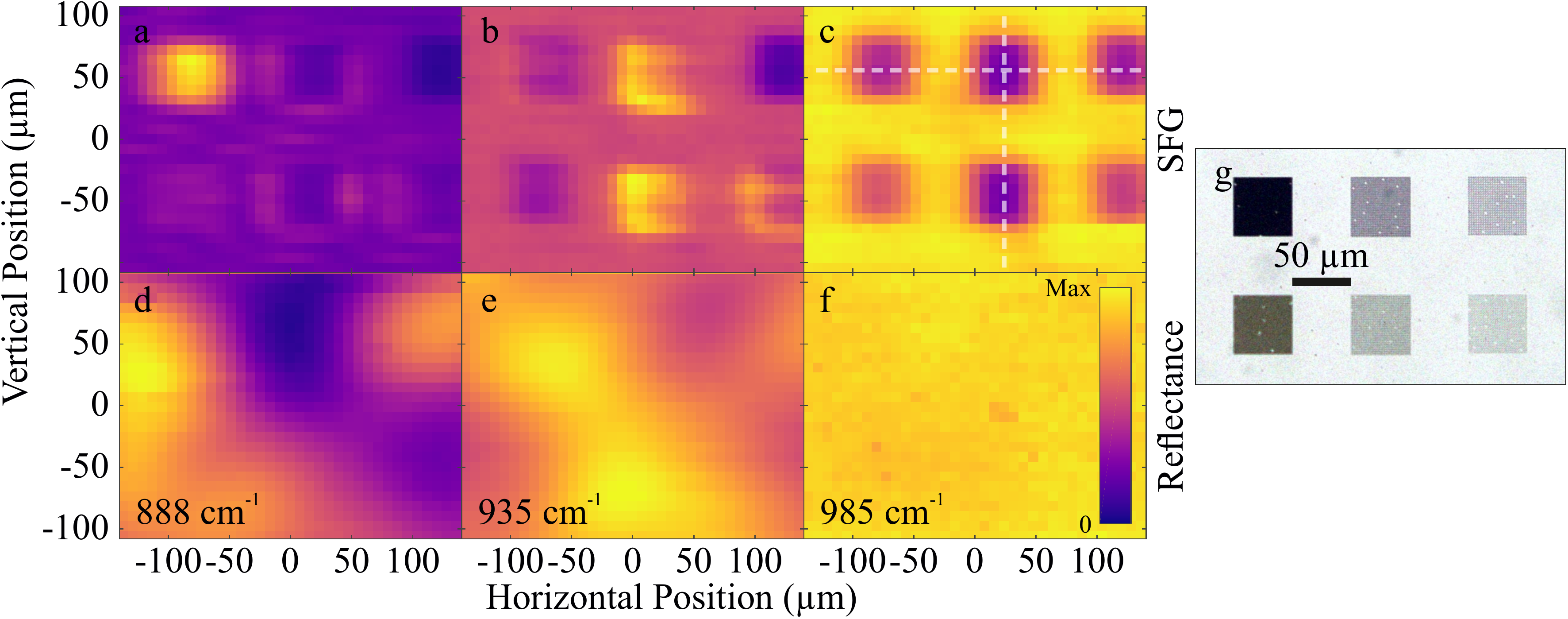}
  \caption{\textbf{Resonant imaging by SFG microscopy.} SFG intensity a-c) and IR reflectance d-f) maps of an area containing six different nanopillar arrays (columns left to right: periods of 700, 1000, 1500~nm; rows: nanopillar diameters of 300~nm (top) and 200~nm (bottom). Maps were acquired at a monopole resonance at $\omega=888~\mathrm{cm}^{-1}$ (a,d), a dipole resonance at $\omega=935~\mathrm{cm}^{-1}$ (b,e), and at the substrate resonance at $\omega=985~\mathrm{cm}^{-1}$ (c,f). The pixel size is 8$\times$8~$\mu$m as set by the step size of the spatial scanning. An optical microscope image of same sample region is shown in g). The SFG images clearly resolve the nanopillar arrays much better than the IR reflectance maps. The geometry-dependence of the nanopillar polariton resonance is well captured in a-b), where only the nanopillar arrays with resonant modes show enhanced SFG intensity. Non-resonant imaging shows no contrast in the reflectance (f), while the suppression of the VIS and SFG effective Fresnel factors results in pronounced nanopillar array contrast in the SFG micro-graph (c). White dashed lines in c) indicate the line scans analyzed in Fig.~\ref{fig3}.
  } 
  \label{fig2}
\end{figure*}

To exploit the spatial resolution accessible by SFG, we focused the beams more tightly, and spatially scanned the sample across an area with multiple arrays of nanopillars with different diameter and period. The resulting SFG microscope images are provided in Fig.~\ref{fig2}a-c) for three different IR wavenumbers. For all SFG images, the nanopillar arrays are well-resolved and can clearly be distinguished from the surrounding  substrate, i.e., we observe a universal SFG imaging contrast. 

Setting the FEL to $\omega=888~\mathrm{cm}^{-1}$ in Fig.~\ref{fig2}a results in drastic SFG enhancement for one array, only. This is consistent with the strong dependence of the monopole resonance frequency on the nanopillar array geometry\cite{Chen2014,Razdolski2016}, such that a steep SFG contrast between the different arrays is well expected in this frequency range. A similar behavior is observed for resonant imaging near the dipolar resonance at  $\omega=935~\mathrm{cm}^{-1}$ shown in Fig.~\ref{fig2}b, however with reduced contrast between the different array geometries. This is also expected, since the dipolar resonance exhibits much reduced spectral dependence on the geometry compared to the monopole mode.\cite{Chen2014} Additionally, the SFG contrast between the nanopillar resonant signal and the substrate is reduced due to the smaller effective field enhancement compared to the monopolar resonance. Similar behavior was also observed in  SHG spectroscopy.\cite{Razdolski2016} Finally, imaging at the substrate Fresnel resonance at $\omega=985~\mathrm{cm}^{-1}$ shows the suppression of SFG signal for all arrays with only mild differences due to the different VIS and SFG scattering efficiencies.

The simultaneously acquired IR reflectance maps in Fig.~\ref{fig2}d-f) exhibit significantly reduced spatial resolution such that the individual nanopillar arrays are barely resolved, in agreement with the estimated size of the FEL focus of $\approx80~\mu\mathrm{m}$ (full-width-at-half-maximum). The reflectance images do show appreciable contrast for the polariton resonances in Fig.~\ref{fig2}d,e). However, no imaging contrast is detectable in the reflectance image at the substrate SFG resonance $\omega=985~\mathrm{cm}^{-1}$. Outside the SiC Reststrahlen band, where no polariton resonances are supported, the subdiffractional size of the nanostructure prevents sizable IR light scattering or localized absorption.

\begin{figure}[ht!]
\includegraphics[width=0.4\linewidth]{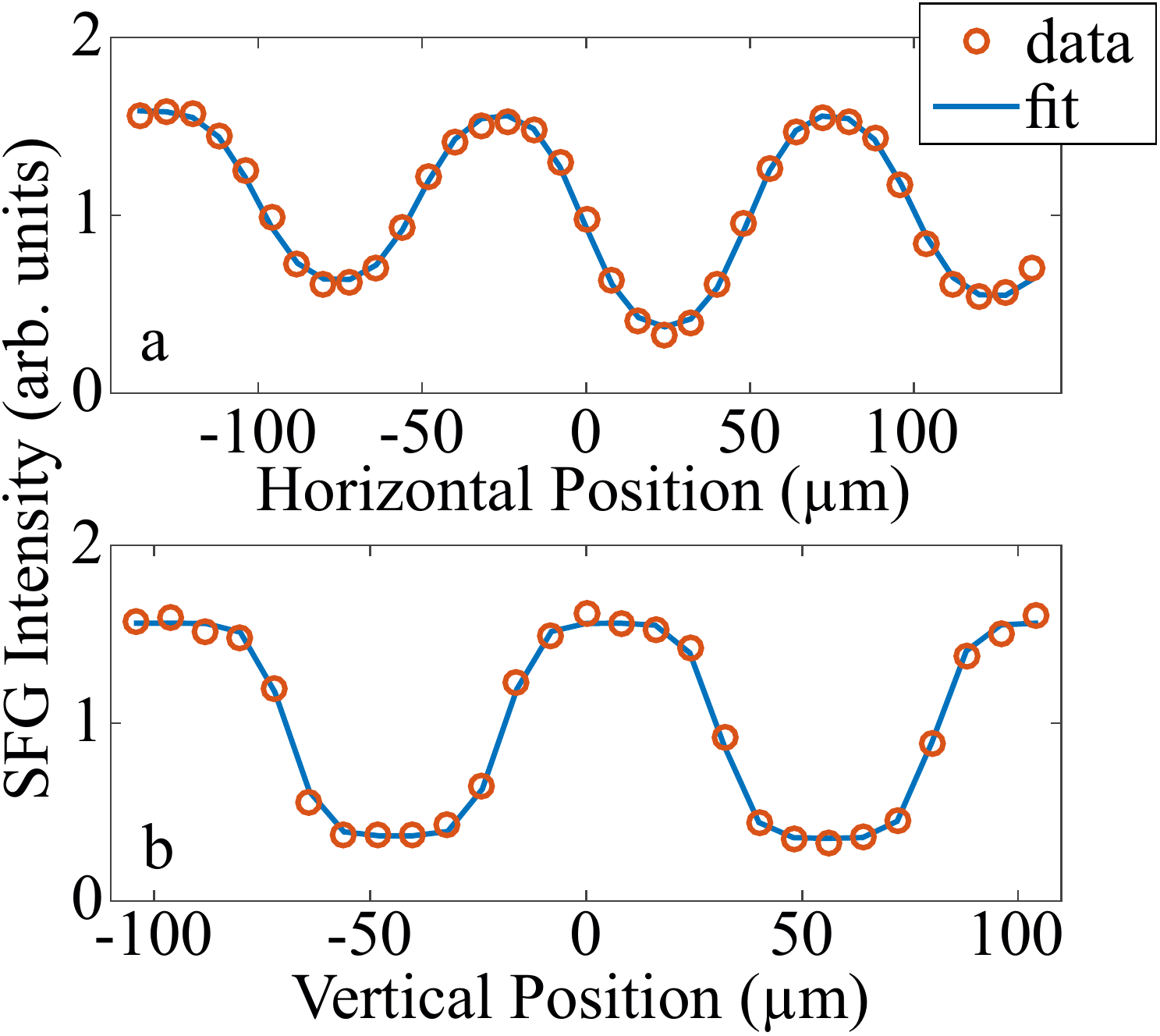}
  \caption{\textbf{Spatial resolution of SFG microscopy.} a) horizontal  and b) vertical line scans (red circles) extracted from the SFG image in Fig.~\ref{fig2}c. The blue lines show the multi-error function fit that yields standard deviations of the Gaussian broadening of 11 and 6~$\mu$m for the horizontal and vertical direction, respectively.} 
  \label{fig3}
\end{figure}

In order to estimate the spatial resolution achieved in our experiments, we analyzed cuts through the SFG images in Fig.~\ref{fig3}. As an example, we there show the SFG signal along the horizontal (Fig.~\ref{fig3}a) and vertical (Fig.~\ref{fig3}b) white-dashed lines in Fig.~\ref{fig2}c. The experimental data were fitted with a sum of error functions representing all the edges between arrays and substrate. The Gaussian broadening globally extracted for each line scan yielded a standard deviation of 11~$\mu$m horizontally and 6~$\mu$m vertically, corresponding to a Rayleigh limited imaging resolution\cite{Haynes2002} of 32 and 18~$\mu$m, respectively. 

Closer inspection of the SFG images in Fig.~\ref{fig2}a and b reveals some inhomogeneities of the SFG signal within and around some of the arrays. Enhancement of SFG signal at the edges of the arrays is clearly observed for the center column in Fig.~\ref{fig2}a, indicating a peculiar behavior of the nonlinear signal when the array symmetry is broken. Similar effects have been observed for electronic resonances in monolayer flakes of van-der-Waals crystals,\cite{Yin2014} and were ascribed to modified electronic structure at the edges of the 2D material. In analogy, our observation here suggests a modified polaritonic mode structure at the edge of the nanopillar array. This would be consistent with the pronounced mode coupling for the monopolar resonances,\cite{Gubbin2017} that is responsible for the strong geometry dependence of the resonant frequency.

\begin{figure*}[ht]
\includegraphics[width=1\linewidth]{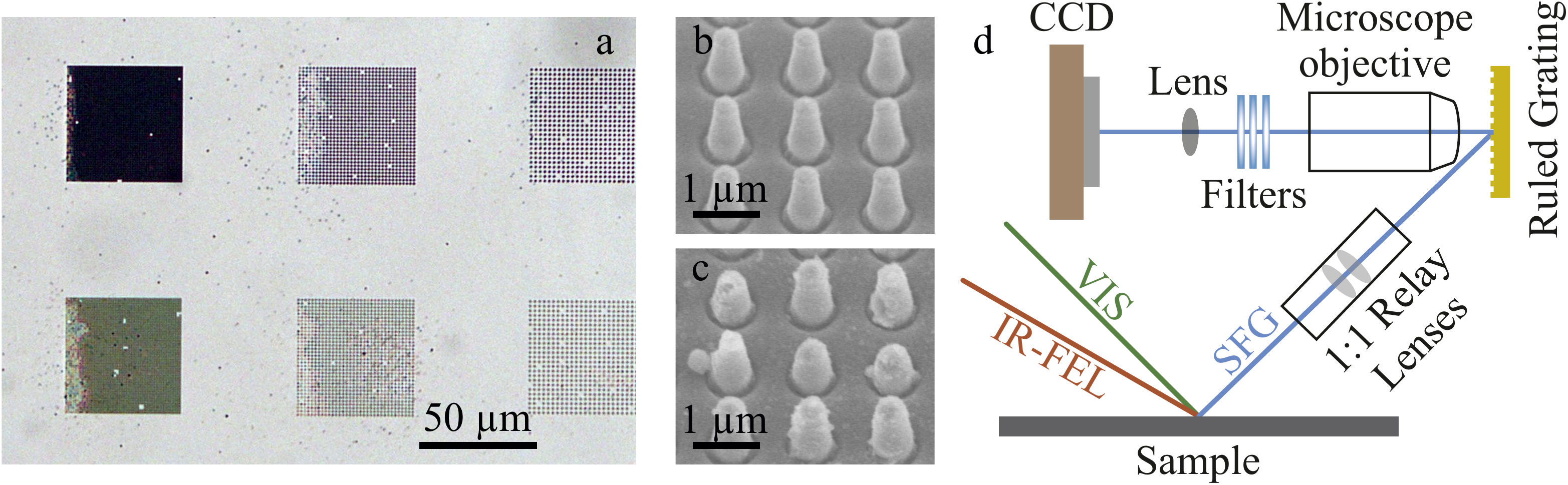}
  \caption{\textbf{Laser-induced sample damage and concept of FEL-based wide-field SFG microscopy.} a) Optical microscope image of the investigated sample area after the SFG experiment. Several patches exhibit some modification indicative of laser-induced damage during the SFG experiment. Despite being irregular, damage is predominantly observed on the left edge of the respective arrays. b-c) Electron microscope images of b) intact and c) laser-damaged nanopillars. The well-localized damage suggests self-saturating resonant heating and partial melting of the nanopillars as the main damage mechanism. d) Schematic of FEL-based wide-field SFG microscopy where tight focusing is avoided thereby significantly reducing the probability for sample damage. This scheme was adapted from previous work on mid-IR wide-field SFG microscopy,\cite{Hoffmann2002} which successfully resolved the problem of image distortion under oblique incidence.}
  \label{fig4}
\end{figure*}

Additionally, we also observed some inhomogeneities in the SFG signal at the dipolar resonance in Fig.~\ref{fig2}b. Here, the peculiar patterns suggest SFG enhancement at the left side for some of the arrays. However, further analysis of the sample revealed some highly localized laser-induced sample damage, see Fig.~\ref{fig4}a for an optical microscope image that was taken after the SFG experiments. There, a clear modification of the sample, particularly on the left side of the arrays is visible. In order to understand the damage mechanism, we took electron microscope images of a damaged sample, see Fig.~\ref{fig4}b-c). The images suggest a very localized heating and partial melting/re-crystallization of the nanopillars. We expect this process to occur for resonant excitation of localized polariton modes. Thereby, the damage process is naturally self-saturating since the heating and melting will result in a loss of the polariton resonance, and, in consequence, much reduced localized absorption preventing further heating. 

We repeated the experiment on similar nanopillar samples with about ten-fold reduced fluences of either the FEL or the VIS laser beams. However, we still observed very similar damage patterns. This supports the interpretation of a self-saturating damage, such that even a ten-fold increase of the fluence, as in the first experiments compared to the later test-measurements, did not magnify the laser-induced damage. We note, though, that we did not observe damage formation when only illuminating with either the FEL or the VIS beams.

Since SFG is a nonlinear process with small photon yield in general,\cite{Shen2016} it requires high power excitation lasers to generate SFG beyond the detection limit. Thus, laser-induced damage is common problem in SFG,\cite{Hoffmann2002} in particular if tight laser foci are required as in scanning probe SFG microscopy.\cite{Raghunathan2011} This problem is amplified for low repetition-rate lasers with high pulse energies, since here the single-shot SFG signal should be maximized to avoid extremely long image acquisition times.\cite{Hoffmann2002} 

Here, wide-field SFG microscopy\cite{Florsheimer1999,Hoffmann2002,Nakai2009,Cimatu2006} as schematically shown in Fig.~\ref{fig4}d presents itself as a complementary approach that solves most of these problems. Most importantly, the spatial resolution in the wide-field approach is no longer determined by the focal sizes of the excitation beams but instead by the resolving power of the imaging optics.\cite{Florsheimer1999} This removes the need to focus the beams tightly, but instead the focal sizes are typically adapted to the field-of-view of the imaging system, leading to a drastic reduction of the laser fluence.\cite{Hoffmann2002} Initial difficulties with image distortions under oblique incidence were successfully resolved using relay imaging onto an appropriate ruled grating.\cite{Hoffmann2002} For non-stationary sample systems, the wide-field approach avoids drift-induced image distortions, and instead potentially allows to follow the evolution of the sample.

With several successful implementations of wide-field SFG microscopy in the mid-IR\cite{Hoffmann2002,Shen2016,Nakai2009,Florsheimer1999,Cimatu2006} and our current work, we expect the realization to LWIR-VIS SFG wide-field microscopy to be straight-forward. Here, we have extended the accessible IR wavelength range of IR-VIS SFG microscopy to beyond 10~$\mu$m with the only fundamental long-wavelength limit being the ability to spectrally separate the SFG signal from the VIS up-conversion laser. Considering that the spatial resolution is theoretically limited solely by the SFG wavelength, this also means LWIR-VIS SFG microscopy should provide imaging capabilities far below the LWIR diffraction limit. Specifically, for SFG wavelength of $500$~nm and IR wavelength of 10~$\mu$m, the Abbe limit of the imaging resolution would be $\lambda_{IR}$/40.

SFG is an even-order nonlinear technique that is typically used to probe the surface vibrations of materials with inversion symmetry in the bulk.\cite{Shen2016} In contrast here, due to the broken inversion symmetry of SiC, the SFG signal emerges from the bulk\cite{Liu2008} and thereby reports on the volume-integrated IR field enhancement of the localized modes inside the nanostructures. In fact, many SPhP materials (most III-V and II-VI semiconductors)\cite{Caldwell2014a} are inversion-broken such that our technique could be employed with ease. For inversion-symmetric SPhP materials, such as bulk hexagonal boron nitride \cite{Dai2014,Caldwell2014} or $\alpha$-Al$_2$O$_3$,\cite{Caldwell2014a} the signals would be significantly weaker and report on the surface fields rather than the volume-confined modes, similar to VIS SHG in plasmonic nanostructures.\cite{Simon1974,Kauranen2012} Additionally, for these systems the bulk-allowed 3$^{rd}$ order process of degenerate IR-IR-VIS SFG\cite{Bonn2001,Cho2009,Grechko2018} is expected with similar magnitude, and could, due to the different wavelength of the emitted SFG signal, be detected independently.\cite{Bonn2001} Therefore, the SFG microscopy approach could also be transferred to study inversion-symmetric materials, where, in principle, both surface and volume-integrated field enhancements could be probed independently.

We also note that SFG microscopy, in contrast to IR s-SNOM, is ideally suited to study buried structures.\cite{Nakai2009} Since SFG preserves the far-field polarization, it allows for distinguishing the normal-to-surface and in-plane optical field enhancements with high sensitivity. When combined with heterodyne detection\cite{Wang2017b,Thamer2018}, the technique could additionally provide phase contrast. Similarly, the sensitivity could be further improved by tuning the VIS wavelength to a resonant electronic transition,\cite{Raschke2002} e.g. to the band gap of the semiconductor under study. Such double-resonant SFG microscopy would provide drastically enhanced signal levels and additional imaging contrast due to the spatial variations of the electronic resonance. For heterogeneous materials, such as semiconductor superlattices, this approach could also serve as a contrast mechanism between the individual constituents.\cite{Ratchford2018,Caldwell2014a} Finally, we expect LWIR-VIS SFG microscopy introduced here to also be  implemented for studies for biological and chemical systems, for instance for probing the spatial heterogeneity of transient low-frequency modes of transient species at electro-chemical interfaces.\cite{Tadjeddine1996,Baldelli1999,Tong2017}

In summary, we have demonstrated scanning-probe LWIR-VIS SFG microscopy of surface phonon polariton resonances in subdiffractional nanostructures. We observed large contrast between the substrate and the nanophotonic structures, both in general for substrate-resonant SFG, as well es specifically when imaging at the polariton resonant frequencies. We achieved a Rayleigh-limited spatial resolution of 18 and 32~$\mu$m along the vertical and horizontal direction, respectively, limited by the focal size of the VIS beam. The tight focusing also led to very localized, self-saturating laser-induced damage to the nanopillars that we ascribed to localized heating through resonant excitation of the polariton modes. We proposed wide-field SFG microscopy as a means to solve this problem and also approach the theoretical spatial resolution given by the diffraction limit of the SFG wavelength. Finally, we discussed the emerging opportunities of LWIR-VIS SFG microscopy, such as IR super-resolution imaging of buried nanostructures or spatial heterogeneities at electro-chemical interfaces.

\begin{acknowledgement}

The authors wish to thank R.J.K. Campen (FHI Berlin) for helpful discussions and W. Frandsen (FHI Berlin) for the electron microscopy. This study was supported by the European Research Council (ERC) under the European Union's Horizon 2020 research and innovation program (grant agreement no 772286). A.G and J.D.C. were supported by the Office of Naval Research through the U.S. Naval Research Laboratory.

\end{acknowledgement}

\begin{suppinfo}
A theoretical description of the 4H-SiC substrate SFG response is available in the Supporting Information.

\end{suppinfo}

\providecommand{\latin}[1]{#1}
\makeatletter
\providecommand{\doi}
  {\begingroup\let\do\@makeother\dospecials
  \catcode`\{=1 \catcode`\}=2 \doi@aux}
\providecommand{\doi@aux}[1]{\endgroup\texttt{#1}}
\makeatother
\providecommand*\mcitethebibliography{\thebibliography}
\csname @ifundefined\endcsname{endmcitethebibliography}
  {\let\endmcitethebibliography\endthebibliography}{}

\end{document}